\documentclass{article}
\usepackage{styfile}
\usepackage{amsmath}
\usepackage{amsfonts}
\usepackage{amssymb}
\usepackage{amsthm}
\usepackage{algorithm,algorithmic}
\usepackage[english]{babel}
\usepackage{mathtools}
\usepackage[binary-units=true]{siunitx}
\usepackage{caption}
\usepackage{graphicx}
\usepackage{hyperref}

\newcommand{\mat}[1]{\ensuremath{{\mathbf{\MakeUppercase{#1}}}}}
\makeatletter
\renewcommand{\vec}[1]{%
	\ifcat\relax\noexpand#1%
	\ensuremath{\boldsymbol{\lowercase{#1}}}%
	\else
	\ensuremath{\mathbf{\lowercase{#1}}}%
	\fi
}
\makeatother
\newcommand{\sumlim}[2]{\ensuremath{\sum\limits_{#1}^{#2}}}

\newcommand{\transpose}[1]{\ensuremath{{#1}^{\textsc{t}}}}
\newcommand{\inverse}[1]{\ensuremath{{#1}^{-1}}}
\newcommand{\R}{\ensuremath{\mathbb{R}}}
\newcommand{\norm}[1]{\left|\left|#1\right|\right|}
\newcommand{\argmin}[1]{\ensuremath{\underset{#1}{\text{argmin }}}}
\newcommand{\sign}[1]{\ensuremath{\text{sign}\!\left(#1\right)}}
\floatname{algorithm}{Algorithm}
\newcommand{\KwIn}[1]{\textbf{Input:} #1}
\newcommand{\KwOut}[1]{\textbf{Output:} #1}

\title{RIEMANNIAN GEOMETRY-BASED DECODING OF THE DIRECTIONAL FOCUS OF AUDITORY ATTENTION USING EEG}
\name{Simon Geirnaert$^{\star \dagger}$ \qquad Tom Francart$^{\dagger}$ \qquad Alexander Bertrand$^{\star}$\thanks{\noindent This research is funded by an Aspirant Grant from the Research Foundation - Flanders (FWO) (for S. Geirnaert), the KU Leuven Special Research Fund C14/16/057, FWO project nr. G0A4918N, the European Research Council (ERC) under the European Union’s Horizon 2020 research and innovation programme (grant agreement No 802895 and grant agreement No 637424), and the Flemish Government (AI Research Program). The scientific responsibility is assumed by its authors.}
\thanks{\noindent S. Geirnaert and A. Bertrand are also affiliated with Leuven.AI - KU Leuven Institute for AI, Belgium.}}

\address{$^{\star}$KU Leuven, Department of Electrical Engineering (ESAT),\\ STADIUS Center for Dynamical Systems, Signal Processing, and Data Analytics, Belgium  \\
	$^{\dagger}$KU Leuven, Department of Neurosciences, ExpORL, Belgium}

\begin{document}

	\maketitle
	
	\begin{abstract}
		Auditory attention decoding (AAD) algorithms decode the auditory attention from electroencephalography (EEG) signals that capture the listener's neural activity. Such AAD methods are believed to be an important ingredient towards so-called neuro-steered assistive hearing devices. For example, traditional AAD decoders allow detecting to which of multiple speakers a listener is attending to by reconstructing the amplitude envelope of the attended speech signal from the EEG signals. Recently, an alternative paradigm to this stimulus reconstruction approach was proposed, in which the directional focus of auditory attention is determined instead, solely based on the EEG, using common spatial pattern filters (CSP). Here, we propose Riemannian geometry-based classification (RGC) as an alternative for this CSP approach, in which the covariance matrix of a new EEG segment is directly classified while taking its Riemannian structure into account. While the proposed RGC method performs similarly to the CSP method for short decision lengths (i.e., the amount of EEG samples used to make a decision), we show that it significantly outperforms it for longer decision window lengths.
	\end{abstract}
	
	\section{Introduction}
	\label{sec:intro}
	
	Previous research has shown that it is possible to decode the auditory attention from brain activity measured by electroencephalography (EEG) sensors~\cite{osullivan2014attentional,geirnaert2020fast,geirnaert2020eegbased}. These \emph{auditory attention decoding} (AAD) algorithms fill an important gap in the design of assistive hearing devices (e.g., cochlear implants or hearing aids), as they inform classical speaker separation and noise reduction algorithms about the speaker a user wants to attend to in a multi-speaker scenario. As such, AAD algorithms constitute a fundamental building block of neuro-steered hearing devices. 
	
	AAD algorithms traditionally use a stimulus reconstruction approach, in which a spatio-temporal decoder is applied to the EEG to reconstruct the amplitude envelope of the attended speaker~\cite{osullivan2014attentional,geirnaert2020eegbased}. The decoded speech envelope then traditionally shows a higher correlation coefficient with the attended speech envelope than with the unattended speech envelope(s). This approach, however, suffers from low decoding accuracy at high speed, i.e., when using few data to decode the auditory attention~\cite{geirnaert2020eegbased,geirnaert2020interpretable}. As these short decision windows (i.e., the amount of data used to decode the attention) $< \SI{10}{\second}$ are paramount for the practical applicability of AAD algorithms, for example, when the attention is switched between two speakers~\cite{geirnaert2020interpretable}, the stimulus reconstruction approach might be too slow for practical neuro-steered hearing devices or for conducting research experiments that require tracking of attention. Furthermore, this approach requires an error-prone speech separation step, in order to retrieve the individual speech envelopes from the recorded mixture of speech sources~\cite{geirnaert2020eegbased,das2020linear}.
	
	As an alternative paradigm, decoding the directional focus of auditory attention, solely based on the EEG, was proposed in~\cite{geirnaert2020fast}. In this approach, the common spatial pattern (CSP) filtering method is used to discriminate between different angular positions of the attended and unattended speaker(s). This CSP approach significantly outperforms the stimulus reconstruction approach on short decision windows. Furthermore, this paradigm does not require a preceding speech separation step. As such, this alternative paradigm improves the practical applicability of neuro-steered hearing devices.
	
	We propose a new AAD algorithm, capitalizing this new paradigm of decoding the directional focus of auditory attention, but replacing the traditional CSP filter method with a so-called \emph{Riemannian geometry classifier} (RGC). This technique has become very popular in the brain-computer interface (BCI) community~\cite{lotte2018review} and outperforms the classical CSP approach in various BCI applications, in particular in motor imagery paradigms~\cite{lotte2018review,barachant2012multiclass,barachant2013classification}. In Section~\ref{sec:methods}, we explain how this RGC can be used to classify the directional focus of auditory attention. In Section~\ref{sec:experiments}, we compare the proposed RGC classifier with the state-of-the-art CSP method and stimulus reconstruction approach. Conclusions are drawn in Section~\ref{sec:conclusion}.
	
	\section{Riemannian geometry-based classification}
	\label{sec:methods}
	
	In recent years, a new class of RGCs has gained a lot of attention in the BCI community~\cite{lotte2018review}. Instead of pre-filtering the EEG using data-driven filters based on the EEG covariance structure (as is the case in CSP filtering~\cite{blankertz2007optimizing}), the EEG covariance matrices are classified directly, as it is assumed that all spatial (and potentially temporal) information concerning different conditions is encoded in these covariance matrices~\cite{barachant2012multiclass,barachant2013classification}. However, covariance matrices are symmetric positive definite (SPD), such that they live on a differentiable Riemannian manifold rather than in a Euclidean space. RGCs take this specific structure into account to improve classification performance. More details about RGCs and their use in BCIs can be found in~\cite{lotte2018review,barachant2012multiclass,barachant2013classification}.
	
	As covariance matrices live on a \emph{differentiable} Riemannian manifold, a tangent space at each point (i.e., covariance matrix) can be computed. Such a tangent space, containing symmetric matrices, is Euclidean, where Euclidean distances between tangent vectors approximate Riemannian distances (i.e., distances between covariance matrices on the Riemannian manifold) of the (projected) covariance matrices. As traditional classifiers rely on Euclidean metrics, which conflict with the Riemannian structure of the manifold on which covariance matrices live, it is preferred to first project all covariance matrices onto the tangent space of a reference matrix. This is the crucial difference with a straightforward direct classification of covariance matrices, which assumes a Euclidean structure of the covariance matrices. In the RGC, the intermediate \emph{tangent space mapping} (TSM) assures that Euclidean metrics are applicable. For the tangent space to be a good local approximation of the Riemannian manifold, where Euclidean distances between tangent vectors closely approximate Riemannian distances between the covariance matrices, a good choice of the reference point of the TSM is the geometric or Riemannian mean.
	
	Let $\{\mat{X}_k,y_k\}_{k=1}^K$ be a training set containing $K$ segments of bandpass filtered EEG data $\mat{X}_k \in \R^{C \times T}$, with $C$ channels and $T$ time samples, and with known labels $y_k \in \{-1,1\}$ (e.g., attending to the left or right speaker). The corresponding covariance matrices are defined as 
	\begin{equation}
		\mat{R}_{k} = \frac{1}{T-1}\mat{X}_k\transpose{\mat{X}}_k \in \R^{C \times C}.
	\end{equation}
	As in~\cite{geirnaert2020fast}, we estimate the covariance matrices using ridge regression, where the regularization hyperparameter is determined automatically using the method proposed in~\cite{ledoit2004well}. This hyperparameter estimation method is considered to be the state-of-the-art in BCI research~\cite{lotte2018review}. 
	
	The geometric or Riemannian mean of these $K$ covariance matrices is then given by the SPD matrix $\mat{R}_{\mathfrak{G}}$ that minimizes the mean squared Riemannian distance from each $\mat{R}_k$ to $\mat{R}_{\mathfrak{G}}$~\cite{barachant2012multiclass}:
	\begin{equation}
		\label{eq:riem-mean}
		\mat{R}_{\mathfrak{G}} = \mathfrak{G}\!\left(\mat{R}_1,\dots,\mat{R}_K\right) = \argmin{\mat{R}  \text{ is SPD}} \sumlim{k = 1}{K}\delta_R^2\!\left(\mat{R}_k,\mat{R}\right),
	\end{equation}
	\noindent
	where $\delta_R\!\left(\mat{R},\mat{S}\right)$ denotes the Riemannian distance between two SPD matrices $\mat{R}$ and $\mat{S}$, which can be computed as~\cite{barachant2012multiclass}:
	\begin{equation}
		\delta_R\!\left(\mat{R},\mat{S}\right) = \norm{\log\!\left(\inverse{\mat{R}}\mat{S}\right)}_F,
	\end{equation}
	with $\log\!\left(\cdot\right)$ the matrix-logarithm. Given a diagonalizable matrix $\mat{A} = \mat{V}\mat{\Lambda}\inverse{\mat{V}}$, the matrix-logarithm of $\mat{A}$ is defined as: 
	\begin{equation}
		\label{eq:log}
		\log\!\left(\mat{A}\right) = \mat{V}\log\!\left(\mat{\Lambda}\right)\inverse{\mat{V}},
	\end{equation}
	with $\log\!\left(\mat{\Lambda}\right)$ a diagonal matrix with diagonal elements $\log(\lambda_{i})$. The Riemannian mean in~\eqref{eq:riem-mean} can only be computed in an iterative way, by iteratively computing the Euclidean mean in the tangent space mapping, or can be approximated using log-euclidean metrics~\cite{congedo2015approximate}:
	\begin{equation}
		\label{eq:log-eucl}
		\mat{R}_{\mathfrak{G}} \approx \exp\!\left(\frac{1}{K}\sumlim{k = 1}{K}\log\!\left(\mat{R}_k\right)\right),
	\end{equation}
	where the matrix-exponential $\exp\!\left(\cdot\right)$ is defined similarly as the matrix-logarithm in~\eqref{eq:log}. We here use the latter estimation method in~\eqref{eq:log-eucl} to efficiently compute the Riemannian mean covariance matrix.
	
	The normalized TSM of the covariance matrix $\mat{R}_k$ onto the tangent space at reference point $\mat{R}_{\mathfrak{G}}$~\eqref{eq:riem-mean} is then equal to~\cite{barachant2012multiclass}:
	\begin{equation}
		\label{eq:tsm}
		\mat{T}_k = \log\!\left(\mat{R}_{\mathfrak{G}}^{-\frac{1}{2}}\mat{R}_k\mat{R}_{\mathfrak{G}}^{-\frac{1}{2}}\right).
	\end{equation}
	\noindent
	The TSM $\mat{T}_k$ is then half-vectorized (i.e., a vectorization over the lower-triangular part only, as it is a symmetric matrix), which leads to the feature vector $\vec{f}_k \in \R^{\frac{C(C+1)}{2}\times 1}$, representing EEG segment $\mat{X}_k$ of the training set. Similarly, for a new test segment $\mat{X}^{(\text{test})}$, the test feature vector can be found by computing the TSM of its covariance matrix using the Riemannian mean $\mat{R}_{\mathfrak{G}}$ over the training set.
	
	The generated feature vectors with the aforementioned method can then be classified using any (Euclidean) classifier, trained with the training set $\{\vec{f}_k,y_k\}_{k=1}^K$. We here choose a support vector machine (SVM) classifier with a linear kernel. Such a classifier works well in high-dimensional feature spaces, which we are dealing with here. Note that combining the TSM with a linear SVM can be interpreted as applying an SVM with a \emph{Riemannian kernel} on the half-vectorized original covariance matrix~\cite{barachant2013classification}. The classification algorithm is summarized in Algorithm~\ref{algo:classification}.
	
	\begin{algorithm}
		\caption{Riemannian geometry-based classification}
		\label{algo:classification}
		\KwIn{Test EEG segment $\mat{X}^{(\text{test})} \in \R^{C\times T}$ and given Riemannian mean $\mat{R}_{\mathfrak{G}}$ over a training set and (linear) SVM classifier $D\!\left(\cdot\right)$}\\
		\KwOut{Class label $y^{(\text{test})}$ (e.g., left or right attended)}
		\begin{algorithmic}[1]
			\STATE Bandpass filter $\mat{X}^{(\text{test})}$ between $\SIrange{12}{30}{\hertz}$
			\STATE Compute a regularized covariance matrix: 
			\[\mat{R}^{(\text{test})} = \frac{1}{T-1}\mat{X}^{(\text{test})}\transpose{\mat{X}^{(\text{test})}} + \delta \mat{I},\]
			with regularization constant $\delta$ 
			\STATE Compute the tangent space mapping onto Riemannian mean $\mat{R}_{\mathfrak{G}}$:
			\[
			\mat{T}^{(\text{test})} = \log\!\left(\mat{R}_{\mathfrak{G}}^{-\frac{1}{2}}\mat{R}^{(\text{test})}\mat{R}_{\mathfrak{G}}^{-\frac{1}{2}}\right)
			\]
			\STATE Compute the feature vector as the half-vectorization $\vec{f}^{(\text{test})} = \text{vech}\!\left(\mat{T}^{(\text{test})}\right)$ of the TSM
			\STATE Classify: $y^{(\text{test})} = \sign{D\left(\vec{f}\right)}$
		\end{algorithmic}
	\end{algorithm}
	
	\section{Experiments and results}
	\label{sec:experiments}
	We compare the proposed RGC method with the CSP method \cite{geirnaert2020fast}, which is the state-of-the-art method for decoding the directional focus of auditory attention. In the CSP method used in~\cite{geirnaert2020fast}, features are generated by applying six spatial filters that maximize discriminability~\cite{blankertz2007optimizing} and are classified with a linear discriminant analysis (LDA) classifier. The state-of-the-art stimulus reconstruction method (canonical correlation analysis (CCA) + LDA), as shown in~\cite{decheveigne2018decoding,geirnaert2020eegbased}, is also added as a reference. For the CCA method, the same preprocessing steps and design choices as in~\cite{geirnaert2020fast} are used.
	
	\subsection{AAD dataset}
	\label{sec:dataset}
	The comparison is performed on a publicly available dataset, which is recorded for the purpose of AAD~\cite{das2019dataset,biesmans2017auditory}. This dataset contains the EEG of $16$ subjects, attending to one of two simultaneously active competing speakers, located at $\pm90^{\circ}$ along the azimuth direction. Per subject, $72$ minutes of data is available. The EEG is recorded using a $C = 64$-channel BioSemi ActiveTwo system. For more details, we refer to~\cite{das2019dataset,biesmans2017auditory}.
	
	\subsection{Design choices}
	\label{sec:design-choices}
	
	\subsubsection{Bandpass filtering}
	\label{sec:prefiltering}
	According to the analysis of the filterband importance in the state-of-the-art CSP approach~\cite{geirnaert2020fast}, the $\beta$-band ($\SIrange{12}{30}{\hertz}$) is the most useful EEG frequency band to decode the directional focus of attention. As such, both for the baseline CSP algorithm, as for the proposed RGC method, the EEG is prefiltered in the $\beta$-band using an 8\textsuperscript{th}-order Butterworth filter and downsampled to $\SI{64}{\hertz}$.
	
	\subsection{Performance evaluation}
	\label{sec:perf-eval}
	The proposed RGC method is tested in a subject-specific way using ten-fold cross-validation. Therefore, the $72$ minutes of EEG data per subject are split into $\SI{60}{\second}$ segments, which are randomly distributed across ten folds. Note that these $\SI{60}{\second}$ segments are normalized by setting the mean per channel to zero, as well as setting the Frobenius norm across all channels to one. The decision window length is defined as the length of the EEG window over which a single AAD decision is made (this usually results in a trade-off between AAD accuracy and decision latency~\cite{geirnaert2020interpretable}). In the case of our RGC framework, the decision window length is defined by the number of samples $T$ over which the covariance matrices are estimated. To evaluate the AAD accuracy for various decision window lengths, all $\SI{60}{\second}$ segments are split into shorter decision windows. The Riemannian mean in~\eqref{eq:riem-mean} and linear SVM are retrained for every decision window length. The significance level for above-chance AAD accuracy is computed based on the inverse binomial distribution~\cite{osullivan2014attentional}. Note that shorter decision window lengths result in more decisions over the test fold, resulting in a lower significance level. A similar ten-fold cross-validation procedure is used for the CSP and CCA method.
	
	Evaluating the AAD accuracy across different decision window lengths is important, for example, in the context of detecting switches in auditory attention. To resolve the traditional trade-off between accuracy and decision window length, the minimal expected switch duration (MESD) metric [s] is used, as proposed in~\cite{geirnaert2020interpretable}. This single-number AAD performance metric quantifies the minimal expected time it takes to switch the gain from one speaker to another, following a switch in attention, based on an optimized stochastic model of a robust (i.e., assuring stable operation above a pre-defined comfort level) attention-steered gain control system.
	
	\subsection{Results}
	\label{sec:results}
\begin{figure}
		\centering
		\includegraphics[width=\linewidth]{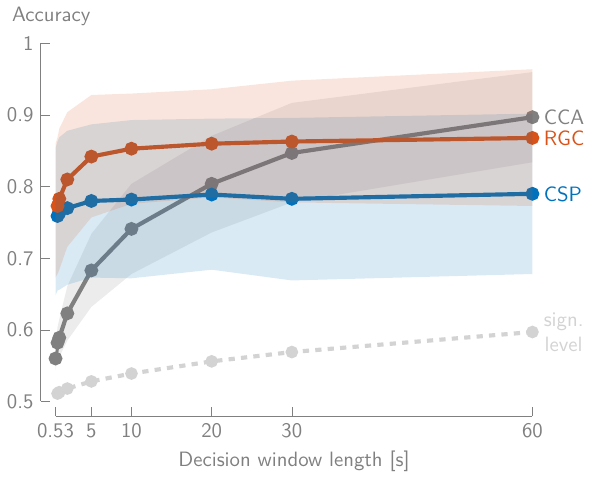}	
		\caption{The mean AAD accuracy across subjects ($\pm$ standard deviation across subjects) shows that the RGC outperforms the CSP approach on almost all decision window lengths, but exhibits a faster decrease in performance on very short decision window lengths, resulting in very similar performances on $\SI{531}{\milli\second}$ decision windows.}
		\label{fig:perf-curve}
	\end{figure}

	Figure~\ref{fig:perf-curve} shows the AAD accuracies as a function of decision window length for the RGC, CSP, and CCA method. Below $\SI{1}{\second}$ decision window lengths (i.e., using $T = 64$ samples at $f_s = \SI{64}{\hertz}$), the RGC and CSP methods have very similar accuracies. Between $\SI{1}{\second}$ and $\SI{5}{\second}$, there is a much faster increase in performance for the RGC method than for the CSP method. This is mostly due to the quickly improving covariance matrix estimation (required for the RGC method) at these shorter decision window lengths. Indeed, as more data become available for increasing decision window lengths to estimate the covariance matrix, less regularization is required, introducing a smaller bias on the estimated covariance matrix. There is no similar effect for the CSP method, as there is no direct covariance matrix estimation involved. Note that, potentially, the RGC method could be improved on these very short decision windows by applying an intelligent dimensionality reduction or feature selection method, which is beyond the scope of this paper. Beyond $\SI{5}{\second}$, the performance levels off in both cases and the RGC method outperforms the CSP method with $\approx 6\%$.
	
	As is also shown in~\cite{geirnaert2020fast}, the traditional stimulus reconstruction method (CCA) outperforms the CSP method for the - less practical - long decision windows $> \SI{20}{\second}$. As the RGC method outperforms the CSP method on almost all decision window lengths, the region in which the CCA method is the best has decreased to the range $> \SI{40}{\second}$. If one would construct an AAD algorithm combining both approaches (RGC + CCA), the envelope would largely, and in the most important regions, be dominated by the RGC method.
	
	The per-subject MESD values are all $< \SI{5}{\second}$ (except for two outliers due to poorer performing subjects with MESDs $> \SI{5}{\second}$, but $< \SI{24}{\second}$), with median MESD $= \SI{2.26}{\second}$ and $[25,75]\%$-quantiles $= [2.13,2.62] \SI{}{\second}$. Note that the MESD values of the CCA method are all above $\SI{5}{\second}$ (due to poor performance at short decision windows, median MESD $ = \SI{16.07}{\second}$). The median MESD of the CSP method is $= \SI{2.34}{\second}$, with $[25,75]\%$-quantiles $= [2.12,2.61] \SI{}{\second}$. For the CSP and RGC method, as there are still relatively high accuracies on the very short decision windows, the optimal trade-off point between AAD accuracy and decision window length is very often located at the shortest decision window lengths. As both methods have very similar accuracies there (see Figure~\ref{fig:perf-curve}), the MESD values are also very similar across both methods, with similar median values. Furthermore, a paired Wilcoxon signed-rank test ($n = 16, p = 0.0627$) shows no significant difference between both methods.
	
	\section{Discussion and conclusion}
	\label{sec:conclusion}
	We have shown that the proposed RGC is capable of outperforming the state-of-the-art CSP method to decode the directional focus of auditory attention by $\approx 6\%$ on most decision window lengths. However, two limitations are to be noted. Firstly, the RGC method performs similarly to the CSP method on very short decision windows (see Figure~\ref{fig:perf-curve}), due to the worse covariance matrix estimation on small sample sizes. As the MESD values indicate that these very short decision windows are most relevant in the context of attention switching, the RGC method achieves a similar overall MESD as the CSP method. Furthermore, this RGC method has due to the TSM in~\eqref{eq:tsm} a higher computational load than applying a simple spatial filter. Both limitations need to be considered for the real-time AAD application in neuro-steered hearing devices.
	
	To conclude, the large increase in AAD accuracy compared to the state-of-the-art CSP method makes the proposed method a good candidate to decode the auditory attention, given that it also outperforms the stimulus reconstruction (i.e., CCA) approach for decision window lengths below $\SI{40}{\second}$. This makes the RGC-based decoding of the directional focus of auditory attention one of the best AAD methods to date.
	
	\bibliographystyle{IEEEtran}
	\bibliography{biblio}
\end{document}